# La óptica de imágenes en la extensión universitaria de Unicamp

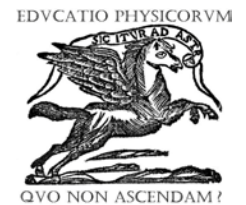

**José Joaquín Lunazzi, Daniel S. F. Magalhães,
Maria Clara Igrejas Amon[1] Rolando Serra Toledo[2]**
[1]*Instituto de Física, P.O.Box 6165, Universidad Estadual de Campinas
UNICAMP, 13083 - 970, Campinas SP, Brasil.*
[2]*Departamento de Física, Instituto Superior Politécnico José Antonio Echeverría -
CUJAE, Ave. 114, 11901, CP 19390, Ciudad de la Habana, Cuba.*

**E-mail:** lunazzi@ifi.unicamp.br

**Resumen**

La enseñanza de la física y en particular de la óptica siempre ha enfrentado problemas de aspecto motivacional principalmente por la pérdida del vínculo con la práctica social y con la vida cotidiana del estudiante. Además, las prácticas de laboratorio han perdido mucho espacio en la enseñanza media por carencias de las escuelas o debido a que tiene apenas como objeto el examen de ingreso a la universidad. El objetivo de este trabajo es mostrar como una actividad de extensión universitaria llamada "Exposición de Holografía" intenta motivar a estudiantes de enseñanza media a través de la experimentación, observación y aplicación de conceptos físicos en la vida cotidiana, despertando sus intereses por la física general.

**Palabras clave:** Física, Exposición de holografía, Extensión universitaria, Óptica.

**Abstract**

The physics teaching particularly the optics teaching always have faced motivational problems, mostly due to the loss of the link between the social practice and the daily life of the student. Moreover, the laboratory practices have lost space in high school attributable to a lack of resources or caused by the fact that the main purpose of the high school is being no more than the access to the University. This work shows how an extension activity called "Holographic Exhibit" tries to motivate high school students by means of experimentation, observation and the application of physics concepts in daily life, awakening their interests to the general physics.

## I. INTRODUCCIÓN

La Universidad es una institución social que tiene como misión transformar la sociedad. Los procesos fundamentales que se llevan a cabo en la misma son [1]:
   Docencia.- Relacionado con la preservación de la cultura.
      Investigación.- Relacionado con la creación de cultura.
      Extensión.- Relacionado con la promoción de la cultura.

La función transformadora debe ser entendida en el sentido de mejorar y perfeccionar la sociedad. Uno de los retos más importantes de la Universidad del siglo XXI es lograr la integración armónica de estos tres procesos dirigidos a la solución de necesidades sociales [2, 3]. La importancia de la función social de la Extensión Universitaria se ha venido analizando en importantes eventos internacionales sobre pedagogía de la Educación Superior [4, 5, 6, 7, 8]. Uno de los objetivos principales de nuestro trabajo es demostrar que es posible realizar un trabajo de Extensión Universitaria a partir de los resultados alcanzados en el trabajo docente y de investigación en el Instituto de Física de la Universidad Estadual de Campinas (UNICAMP).

Entre las principales dificultades del proceso de enseñanza-aprendizaje de la Física en la enseñanza media podemos señalar las siguientes [7]:
- Poca motivación.
- Poca comprensión de los principios, leyes y conceptos que se estudian y tratamiento insuficiente de algunos temas como la Óptica.
- Habilidades experimentales y de observación muy limitadas.
- Muy poca utilización de medios de enseñanza.
- No se analizan al nivel requerido las aplicaciones en la práctica profesional y en la vida cotidiana.

*J. Lunazzi , D. Magalhães , M.C. Igrejas, R. Serra*

La Exposición de Holografía es el nombre que ha tenido la actividad de Extensión Universitaria desarrollada durante varios años por el Profesor J. J. Lunazzi, del Laboratorio de Óptica del Instituto de Física de la UNICAMP, con la participación de alumnos y monitores que han colaborado en el desarrollo de esta actividad, que ha estado dirigida fundamentalmente a estudiantes de la enseñanza media, como una contribución para modificar esta situación [10].

Para fundamentar pedagógicamente la concepción de utilización de la Exposición de Holografía, partiremos del análisis de los principios pedagógicos que se derivan de las leyes generales de la pedagogía [9] y que sirven de base a esta investigación. De las diferentes clasificaciones de principios pedagógicos que aparecen en la literatura, asumiremos la planteada por Addine F. [11], por estar enfocada a la dirección del proceso pedagógico. Ellos son:

- Principio del carácter científico del proceso pedagógico.
- Principio de la vinculación de la educación con la vida, con el medio social y el trabajo, en el proceso de educación de la personalidad.
- Principio de la unidad de lo instructivo, lo educativo y desarrollador, en el proceso de educación de la personalidad.
- Principio de la unidad de lo afectivo y lo cognitivo, en el proceso de educación de la personalidad.
- Principio del carácter colectivo e individual de la educación y el respeto a la personalidad del educando.
- Principio de la unidad entre la actividad, la comunicación y la personalidad.

Analicemos a continuación como estos principios han servido de sustento y de guía para la concepción de utilización de la Exposición de Holografía como una actividad de Extensión Universitaria.

El principio del carácter científico del proceso pedagógico nos indica que debe ser estructurado teniendo en cuenta los avances continuos de la ciencia y la técnica. En los últimos años se ha planteado en diversas conferencias, artículos científicos y en reuniones de importantes organismos internacionales la necesidad de implementar nuevas tecnologías en las universidades que incluyen la introducción y utilización de nuevos medios de enseñanza dentro del perfeccionamiento de la Educación Superior en el mundo [12, 13, 14]. Es en esta dirección que se estructuró nuestra propuesta de utilización de los hologramas y otros experimentos y aplicaciones de la óptica en la formación de imágenes.

El principio de la vinculación de la educación con la vida, con el medio social y el trabajo en el proceso de educación de la personalidad es muy importante en nuestra concepción, al proporcionar el vínculo de los estudiantes con experimentos y fenómenos de la Física en particular de la Óptica y la formación de imágenes presentes en la vida cotidiana, además porque apunta a que la sociedad participa en la educación de todos sus ciudadanos. Este principio también sirve de base a la concepción de considerar los procesos universitarios de Investigación, Docencia y Extensión Universitaria, tradicionalmente no relacionados, como un único proceso integrado donde la actividad de extensión permite la aplicación de los resultados de la investigación y de la actividad docente a la educación social.

El principio de la unidad de lo instructivo, lo educativo y lo desarrollador en el proceso de educación de la personalidad tiene como fundamento la unidad dialéctica existente entre la instrucción y la educación en el proceso de formación y desarrollo de la personalidad. En la concepción de las actividades que integran la Exposición de Holografía se tuvieron en cuenta no solo los conceptos propios de la óptica y la formación de imágenes, sino también los elementos históricos relacionados con la vida y obra de los principales científicos que aportaron al tema y de las diversas aplicaciones en la vida cotidiana de las técnicas holográficas, aspectos relacionados con la formación educativa de los estudiantes.

El principio de la unidad de lo afectivo y lo cognitivo, en el proceso de educación de la personalidad es básico en nuestra propuesta de actividad debido a que los hologramas y el resto de las demostraciones de óptica de imágenes son altamente motivadoras, produciéndose la unidad entre lo afectivo, lo cognitivo y lo educativo. Esta influencia positiva en la motivación que brindan las demostraciones y experimentos con imágenes y sus interesantes aplicaciones en la vida cotidiana, predisponen favorablemente al estudiante y contribuyen significativamente a su motivación por el estudio de la Física.

Si partimos de que la motivación es esencialmente un impulso en la actuación del sujeto y que motivar quiere decir crear interés, estimular el deseo, llamar la atención, despertar la curiosidad, contagiar con entusiasmo y suscitar el gusto como impulso activador [15] llegamos a la conclusión de que nuestra propuesta de Exposición de Holografía es altamente motivante y propiciadora de la unidad entre lo afectivo y lo cognitivo.

Para lograr comprender la importancia de tener en cuenta la dimensión afectiva del aprendizaje desde la infancia hasta la vida adulta se requiere dar la prioridad necesaria a las relaciones, teniendo en cuenta las dimensiones social e interactiva del proceso educacional. Una negación de la dimensión afectiva constituye un serio problema educacional [16]. Este aspecto se tuvo en cuenta en el diseño de la actividad garantizándose el dinamismo y un alto grado de interacción entre los alumnos y entre los alumnos y los profesores.

Los profesores ocupan una posición extremadamente importante en los procesos mentales de sus alumnos [17], por lo que otra dimensión afectiva que podemos resaltar en nuestra propuesta es que propicia las relaciones de los estudiantes con sus profesores que los acompañan en el desarrollo de toda la propuesta mediante la constante interacción prevista en el desarrollo de las actividades.

En la psicología contemporánea se desarrolló y consolidó en la segunda mitad del siglo pasado un enfoque epistemológico originado a partir de la escuela histórica cultural de L. S. Vigotsky y seguidores [18, 19, 20, 21, 22]. Con relación a la importancia de la motivación en el aprendizaje, Vigotsky a partir de reconocer el carácter integral del psiquismo humano, analiza las relaciones existentes entre dos esferas tradicionalmente escindidas en las escuelas psicológicas precedentes: la esfera cognoscitiva y la afectiva. En el primer capítulo de su libro Pensamiento y Lenguaje señala [19]:

*"La primera cuestión que surge cuando hablamos de la relación del pensamiento y el lenguaje con respecto a los restantes aspectos de la conciencia, es el de la vinculación entre la inteligencia y el afecto. Como se sabe, la separación del aspecto intelectual de nuestra conciencia y del aspecto afectivo, volitivo, es uno de los defectos fundamentales y radicales de toda la psicología tradicional". Más adelante señala: "El análisis que divide el todo complejo en unidades... muestra que existe un sistema dinámico de sentido que representa la unidad de los procesos afectivos e intelectuales. Muestra que en toda idea se contiene, reelaborada, una relación afectiva del hombre hacia la realidad, representada en esa idea. Permite descubrir el movimiento directo que va de la necesidad de los impulsos del hombre a la determinada dirección de su pensamiento, y el movimiento contrario, desde la dinámica del pensamiento a la dinámica del comportamiento y la actividad concreta de la persona".*

Paulo Freire construye teoría desde la propia práctica. En alguna de sus últimas obras menciona a Vigotsky, siendo punto de encuentro entre ambos la importancia del rol docente (mediación) en el proceso educativo y los requerimientos de formación del mismo para poder generar desafíos, necesarios en el proceso de enseñanza aprendizaje (zona de desarrollo potencial) [23].

El principio del carácter colectivo e individual de la educación y el respeto a la personalidad del educando está presente en la concepción de realización de nuestra actividad con la participación de un grupo de estudiantes en la conferencia inicial y después la atención mas individualizada al dividirse el grupo en diferentes equipos de trabajo.

El principio de la unidad entre la actividad, la comunicación y la personalidad considera que esta se forma y se desarrolla en la actividad y en el proceso de comunicación [15]. Los medios de enseñanza utilizados en el desarrollo de la Exposición de Holografía favorecen la comunicación que se establece entre el profesor y el alumno, entre los propios alumnos y en general entre todos los participantes de este proceso, influyendo directamente en el desarrollo y educación de la personalidad. Para lograr este objetivo la actividad se estructura de manera que facilite la mejor y más efectiva comunicación posible. El papel de estos medios es establecer los vínculos necesarios entre los niveles sensoriales y racionales del conocimiento, entre lo concreto y el pensamiento abstracto; es así donde pueden ayudar realmente al aprendizaje de los estudiantes, a hacer más comprensibles los conceptos, y abstraerse más fácilmente, a representar en su mente con más claridad aquellas cosas que al profesor son sumamente claras e incuestionables.

Otro aspecto importante que Paulo Freire señala es que *"enseñar exige respeto a la autonomía del ser educando"* [24], así el profesor debe respectar la curiosidad del

*La óptica de imágenes en la extensión universitaria de Unicamp*

alumno, su gusto estético, su inquietud, su lenguaje, dando así libertad para que el alumno piense y construya su conocimiento. En la Exposición de Holografía siempre los estudiantes son estimulados a hacer preguntas, a participar de los experimentos, o sea, constituyen sujetos activos y autónomos del conocimiento.

Otro de los elementos importantes desde el punto de vista pedagógico que se tuvo en cuenta en nuestra propuesta de utilización de la óptica de imágenes es la importancia de las reproducciones visuales [25]:

*"Las reproducciones visuales son las más complejas de todas las reproducciones, pero además, la más importante en la enseñanza, debido al valor que tiene la percepción visual, tanto para la recepción de información como para su retención. Con los medios de enseñanza se aprovechan en mayor grado las potencialidades de nuestros órganos sensoriales. El 83 % de lo que el hombre aprende le llega a través del sentido visual".*

## II. MATERIALES Y MÉTODOS

Para poder diseñar y desarrollar la Exposición de Holografía es preciso definir los objetivos que tendría esta actividad de Extensión Universitaria.

Objetivos generales de la Exposición de Holografía como actividad de Extensión Universitaria:

- Seleccionar contenidos transferibles a situaciones de la vida cotidiana, que favorezcan el aprendizaje colectivo y la interacción grupal.
- Hacer de la actividad un proceso pedagógico vinculado a lo que rodea al estudiante en lo social, lo económico, lo familiar, lo productivo y a la naturaleza.
- Desarrollar habilidades en el alumno para trabajar en grupo, para que aprenda con los otros y de los otros, interactuando cooperativa y solidariamente.
- Preparar a los estudiantes para comprender las problemáticas acuciantes del mundo de hoy, a través de actividades que permitan asimilar los conocimientos científico-técnicos y desarrollar iniciativas.
- Contribuir al incremento de la motivación de los estudiantes de la enseñanza media por el estudio de la física y sus aplicaciones.

En el diseño de las actividades que integrarían esta Exposición de Holografía y en la definición de los métodos que se utilizarían en la misma, es importante precisar los aspectos que se deben tener en cuenta para desarrollar la motivación en los estudiantes:

1. Despertar la curiosidad. La exploración y la curiosidad son motivos activados por lo nuevo y lo desconocido. Es de gran importancia que los aprendizajes tengan un valor significativo. En la medida que los contenidos propuestos puedan resultar cercanos al mundo del estudiante o puedan tener una aplicación práctica real tendrán un mayor valor motivacional.

*J. Lunazzi , D. Magalhães , M.C. Igrejas, R. Serra*

2. Generar sensación de control. Es necesario que el estudiante tenga conciencia de su capacidad para desarrollar los aprendizajes que se le proponen.
3. Promover el sentido de la responsabilidad. El estudiante debe participar de forma responsable en las actividades.
4. Proponer metas con un grado moderado de dificultad. Una tarea excesivamente fácil pierde interés para el estudiante y le conduce al aburrimiento. Una tarea con dificultad excesiva puede hacerle sentirse superado y abandonar.
5. Favorecer el aprendizaje independiente. Es conveniente que el estudiante se enfrente de manera individual a las tareas planteadas.
6. Proporcionar seguridad y apoyo. Es aconsejable que el estudiante sienta la presencia del profesor o del monitor, en caso de encontrar dificultades, que le proporcione el apoyo necesario para resolver las tareas con éxito.
7. Valorar el esfuerzo insistiendo en que los errores son parte del aprendizaje. El estudiante necesita ver recompensado su esfuerzo, por lo que los profesores tienen que atender más al proceso que al resultado.
8. Enseñar a atribuir el éxito a variables controlables (el esfuerzo, la constancia, la ayuda del profesor) en vez de hacerla depender de variables inconsistentes como la suerte o la casualidad.
9. Insistir en lo positivo antes que criticar lo negativo ayudará al estudiante a sentirse competente para la realización de la tarea propuesta y animarlo a intentar mejorar lo que todavía no ha conseguido.
10. Exigir de forma realista y comprensiva. Debemos tener muy en cuenta las posibilidades y capacidades de los estudiantes y exigirles en consecuencia.
11. Intentar ser el mejor ejemplo para ellos. El mejor estímulo será siempre intentar ser un buen modelo de actitud al que puedan imitar.
12. Explicar a los alumnos los objetivos educativos que tenemos previstos para la actividad, principalmente relacionado con la realización de experimentos.
13. Plantearles las actividades de forma lógica y ordenada.
14. Proponerles actividades que les hagan utilizar distintas capacidades para su resolución.
15. Fomentar la comunicación entre los alumnos y las buenas relaciones, realizando tareas de grupo.
16. Aplicar los contenidos y conocimientos adquiridos a situaciones próximas y cercanas para los alumnos.

La Exposición de Holografía constituye el primer módulo de un sistema concebido para el estudio integral de la holografía y sus aplicaciones. Esta propuesta de actividad en la Exposición de Holografía tiene sus categorías específicas, las cuales pueden ser caracterizadas como:
Objetivo General: Planteado en términos de lograr la contribución a la formación de una cultura general integral en los estudiantes de enseñanza media mediante el conocimiento de diferentes tipos de imágenes y sus aplicaciones y el incremento de la motivación por la Física.
Contenidos:
Relacionados con:
- Fundamentos de la óptica de imágenes y sus aplicaciones.
- Fundamentos y aplicaciones de la Holografía.

Métodos:
Se utiliza el método expositivo, de elaboración conjunta y de participación individual en las diferentes etapas de la actividad.
Medios:
- Conferencia inicial con demostraciones sobre diferentes tipos de imágenes.
- Diferentes tipos de hologramas.
- Demostraciones experimentales sobre diferentes tipos de imágenes.

Evaluación: Se realiza a través de instrumentos elaborados como encuestas, criterio de expertos y entrevistas con el objetivo de monitorear el proceso y poder solucionar las deficiencias que se presenten.

La actividad de Exposición de Holografía consta de 4 partes esenciales que describiremos a continuación:
1. Conferencia inicial.
2. Experimentos demostrativos sobre reflexión, refracción, difracción de la luz e imágenes con espejos.
3. Holoproyector y exposición de hologramas.
4. Conferencia final.

**A. Primera Parte: Conferencia inicial**

Duración: 60 minutos
Objetivo general:
Presentar los conceptos básicos de formación de imágenes, reflexión, refracción y difracción y realizar algunas demostraciones con la participación activa de los estudiantes para lograr la motivación inicial de los mismos y los conocimientos mínimos que permitan entender adecuadamente los experimentos y demás demostraciones en el resto de la actividad.
Contenido:
- Presentación de un holograma como elemento motivador inicial.
- Presentación de los conceptos básicos de formación de imágenes, reflexión, refracción y difracción.
- Formación de sombras y radiografía.
- Evolución histórica de la construcción de espejos a partir de los antecedentes presentes en la arqueología americana.
- Aplicaciones de la difracción, holografía y de desarrollos realizados en la UNICAMP como holoimágenes y televisión holográfica.
- Importancia de la visión binocular en el hombre y realización de demostración interactiva con los estudiantes.

- Demostraciones de estereoscopia donde los estudiantes aprecian imágenes con la utilización de anteojos (espejuelos) de dos colores.

La figura 1 muestra la parte de la conferencia inicial donde los alumnos observan imágenes estereoscópicas utilizando anteojos de dos colores repartidos por los monitores.

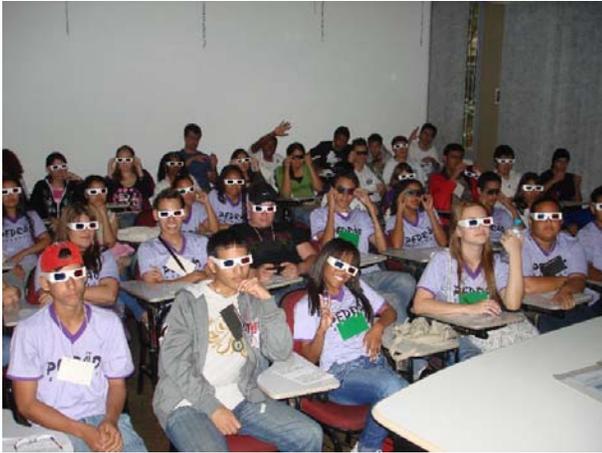

**FIGURA 1.** Alumnos con anteojos de dos colores observando las imágenes estereoscópicas.

**B. Segunda parte: Experimentos demostrativos sobre reflexión, refracción, difracción de la luz e imágenes con espejos**

Duración: 30 minutos (Divididos en tres equipos)
Objetivo general: Realización de experimentos demostrativos con materiales simples y construidos por los monitores sobre los fenómenos presentados en la conferencia inicial, para que los estudiantes puedan apreciar y entender las características fundamentales de los mismos y despertar el interés por la física en particular por la óptica.
Contenido:

- Experimentos de reflexión en piedras pulidas y espejos planos, cóncavos y convexos.
- Experimento demostrativo con la participación individual de los estudiantes de observación de imágenes con un espejo plano especialmente diseñado para ser colocado en la nariz a la altura de los ojos [26]. Los estudiantes experimentan la sensación de estar caminando sobre las nubes, el techo o los árboles y es altamente motivante.
- Experimentos de refracción de la luz y de formación de imágenes con diferente profundidad.
- Demostración del funcionamiento de una cámara fotográfica antigua.
- Experimentos de difracción en redes y en discos compactos.

Las figuras 2 y 3 muestran diferentes momentos de esta segunda parte de la actividad.

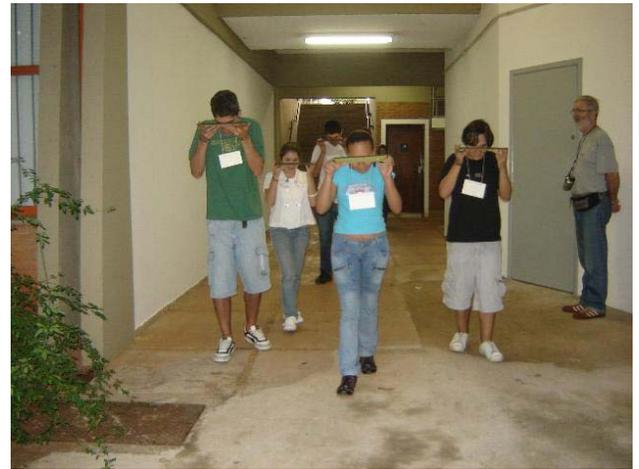

**FIGURA 2.** Alumnos experimentando con espejos.

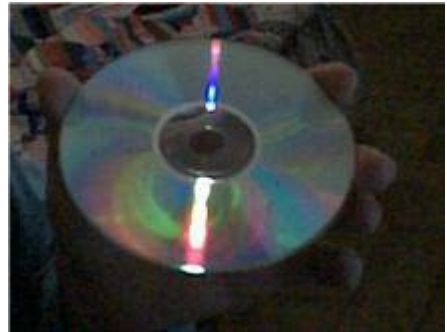

**FIGURA 3.** Difracción en un disco compacto.

**C. Tercera parte: Experimento demostrativo de holoimágenes y exposición de diferentes tipos de hologramas**

Duración: 20 minutos
Objetivo general: Que los alumnos puedan observar diferentes tipos de hologramas reconstruibles con luz blanca y apreciar técnicas de avanzada con la realización de un experimento demostrativo sobre holoimágenes utilizando una pantalla holográfica construida en la UNICAMP.
Contenido:

- Demostración de holoimágenes con la utilización de un holoproyector y una pantalla holográfica.
- Exposición de diferentes tipos de hologramas reconstruibles con luz blanca de objetos inanimados, de personas y de aplicaciones a la medicina. En algunos es posible observar la imagen formada frente de la placa holográfica.

En las figuras 4 y 5 se muestran estudiantes apreciando la exposición de hologramas y uno de los que integra la muestra.

*J. Lunazzi , D. Magalhães , M.C. Igrejas, R. Serra*

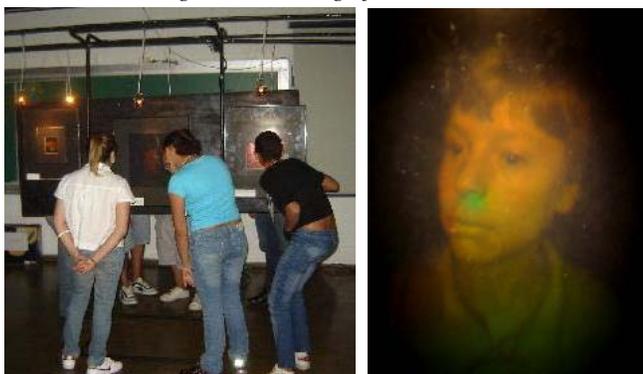

**FIGURAS 4 y 5.** Exposición de diferentes tipos de hologramas.

### D. Cuarta parte: Conferencia final

Objetivo general: Lograr con la participación de un profesor de reconocida experiencia la integración de los temas tratados en la actividad y que los estudiantes conozcan avances recientes en el tema de las holoimágenes y otros aspectos de actualidad relacionados con su formación integral.

Duración: 20 minutos

Contenido:
- Integración de los temas tratados en la actividad.
- Panorámica de avances recientes en la temática de las holoimágenes.
- Aspectos de actualidad de interés relacionados con la formación integral de los estudiantes.
- Distribución de un DVD de elaboración propia que contiene quince videos de experimentos de Física para los profesores responsables de la actividad [27].
- Hacer referencia histórica a las primeras publicaciones sobre el tema [28] y llamar la atención de que desde el año 2002 es que se cuenta con los recursos de la proyección audiovisual digital, que permiten una dinámica mayor en las presentaciones, con actualización simple de figuras, y bajo costo de la realización de imágenes.

## III. RESULTADOS Y DISCUSIONES

Para valorar con mayor precisión la efectividad del trabajo realizado en la Exposición de Holografía, se realizó una encuesta a una muestra de 51 estudiantes de dos de las escuelas participantes en esta actividad. A continuación mostramos la encuesta aplicada, los resultados obtenidos y un análisis de los mismos, donde se evidencia claramente la utilidad de este tipo de actividad de Extensión Universitaria como contribución a la formación de una cultura general integral en los estudiantes y en el incremento de la motivación por la Física y sus aplicaciones. Encuesta aplicada:

1. ¿Qué piensa usted de la actividad en la Exposición de Holografía de la Unicamp?
a) Me gustó mucho   b) Me gustó   c) Regular   d) No me gustó

2. ¿Usted sabía cómo se forman las imágenes?
a) Si, lo aprendí en la escuela   b) Si, lo aprendí solo   c) Tenía alguna idea   d) No, aprendí en la actividad   e) No y continúo sin saber

3. ¿Usted sabía que era la Holografía?
a) Sí   b) Tenía alguna idea   c) Leí alguna cosa sobre eso   d) No, aprendí en la actividad   e) No y continúo sin saber

4. ¿Usted ya había visto imágenes estereoscópicas (vistas con anteojos de dos colores)?
a) Sí, varias veces.   b) Sí, algunas veces.   c) Nunca las había visto.

5. ¿Dé una nota para la conferencia inicial de 0 a 10?

6. Dé una nota de 0 a 10 para cada uno de los experimentos:
- Experimentos de reflexión y difracción
- Experimentos de refracción
- Holoproyector y hologramas
- Espejos "La Nube"

7. Con relación al trabajo de los monitores:
a) Muy bueno   b) Bueno   c) Regular   d) Malo

8. ¿Usted quisiera regresar a los próximos módulos?
a) Sí   b) No   c) Quizás

9. ¿Qué fue lo que le llamó más la atención en la actividad?

10. ¿Qué fue lo que más le gustó?

11. ¿Existió algo que a usted no le gustó?

Usted puede escribir cualquier otro comentario o sugerencia que considere necesario:

A continuación mostraremos los principales resultados de las encuestas aplicadas

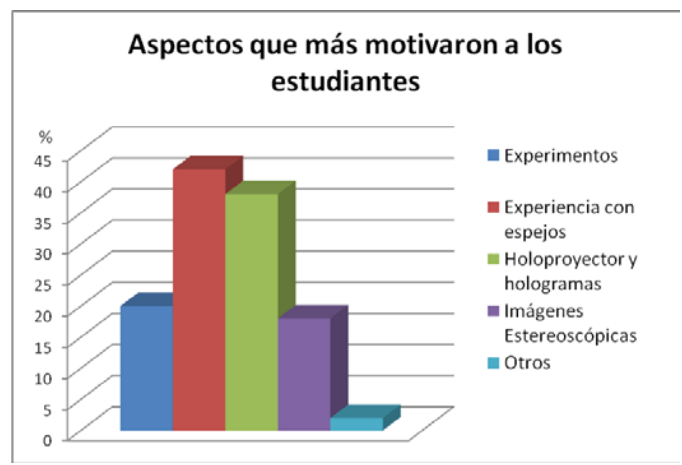

**FIGURA 6.** Resultado de la encuesta acerca de los aspectos que más motivaron a los estudiantes en la Exposición de Holografía.

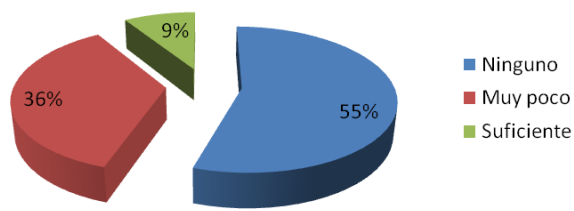

**FIGURA 7.** Resultado que muestra los conocimientos previos de los estudiantes sobre la temática de formación de imágenes.

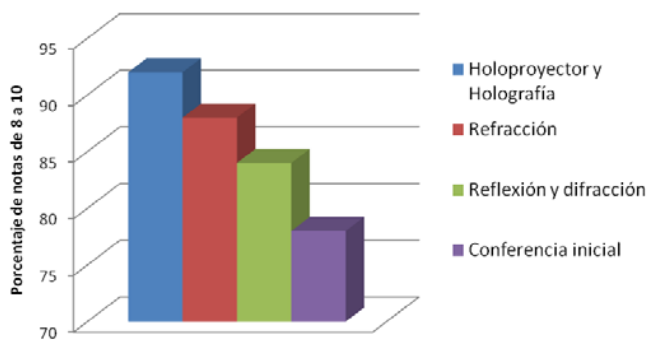

**FIGURA 8.** Resultado de la evaluación realizada por los estudiantes sobre la calidad de las actividades. En el gráfico analizamos solamente el porcentage de notas de 8 a 10 en una escala de 1 a 10.

Análisis de los resultados de las encuestas:
- Al 100% de los estudiantes le gustó la actividad.
- Solo un 9% de los estudiantes manifestaron tener un conocimiento previo suficiente de los temas tratados en la actividad.
- La gran mayoría de los estudiantes evaluó la calidad de las actividades realizadas con notas sobresalientes entre 8 y 10, siendo la conferencia inicial la de más bajos resultados en esta categoría con el 78%.
- El 98% de los estudiantes consideró adecuado el trabajo de los monitores.
- Lo que más motivó a los estudiantes fue el experimento interactivo con los espejos (42%) y los hologramas (38%), seguido por el resto de los experimentos (20%) y las imágenes estereoscópicas (18%).
- Ningún estudiante expresó la decisión de no regresar para próximos eventos convocados.

El 15% de los estudiantes expresaron algunos criterios sobre aspectos que no le gustaron y algunas sugerencias

*La óptica de imágenes en la extensión universitaria de Unicamp*

para perfeccionar la actividad entre las cuales podemos señalar: debe tenerse más cuidado en algunos términos utilizados que no fueron bien comprendidos, debe revisarse la forma en que fueron expuestos algunos temas de la conferencia inicial y deben revisarse algunas de las demostraciones de los experimentos de reflexión, refracción y difracción para que sean mejor comprendidas y mas motivantes.

## IV. CONCLUSIONES

En el trabajo se realizó la fundamentación pedagógica y psicológica de la Exposición de Holografía como actividad de Extensión Universitaria realizada por el Grupo de Holoimágenes del Laboratorio de Óptica del Instituto de Física de la Universidad de Campinas con estudiantes de la enseñanza media.

Entre las principales deficiencias del proceso de enseñanza-aprendizaje de la física en la escuela media están la falta de motivación e interés de los estudiantes, la poca realización de actividades prácticas y el vínculo insuficiente de los temas tratados con la vida cotidiana. En este sentido, la actividad de Exposición de Holografía cumplimentó los objetivos propuestos al lograrse un conocimiento en los estudiantes de enseñanza media sobre la formación de imágenes y sus aplicaciones en la vida cotidiana, un incremento considerable en la motivación por el estudio de la Física y en especial por la Óptica y el desarrollo de habilidades para el trabajo en grupo mediante una interacción cooperativa y solidaria.

## REFERENCIAS

[1] Díaz, T., La Extensión: *Un proceso formativo de la Universidad. Su relación con otros procesos*, Conferencia Magistral del VI Taller Internacional de Extensión Universitaria, Cuba, (2001).
[2] Bricall, J., *Cambios previsibles en la Enseñanza Superior*, Conferencia Especial del evento Universidad 2004, Cuba, (2004).
[3] De la Fuente, J., *Los retos de la Educación Superior Contemporánea*, Conferencia Especial del evento Universidad 2004, Cuba, (2004).
[4] González, G., *La Universidad ante el reto de la formación cultural integral. Aproximación al programa nacional de Extensión Universitaria*, Memorias del evento Universidad 2002, Cuba, (2002).
[5] Grandinetti, V., *Aportes para una Universidad del siglo XXI*, Memorias del evento Universidad 2002, Cuba, (2002).
[6] Nuguer, L., *La función social de la Universidad y la Extensión*, Memorias del evento Universidad 2002, Cuba, (2002).
[7] Serra, R. *La utilización del holograma como medio de enseñanza y de educación social en Cuba a través del vínculo Investigación – Docencia – Extensión Universitaria,* Tesis Doctoral, Cuba, (2004).

[8] Serra, R., *La utilización del holograma en docencia y museología en Cuba*, Memorias XLIX Congreso Nacional de Física de México, Universidad Autónoma de San Luis de Potosí, 16 – 20 de Octubre, (2006).

[9] Calzado, D., *La Ley de la Unidad de la Instrucción y la Educación*. Didáctica Teórica y Práctica, Capítulo 2., 18-29, Cuba, (2003).

[10] Lunazzi, J. J., Magalhães D. S. F. and Rivera N. I. R., *Didactical holographic exhibit including holotv (holographic television)*, Riao/Optilas'07, Proceeding **992**, 210-215, (2007).

[11] Addine, F., *Principios para la dirección del proceso pedagógico*, Compendio de Pedagogía, Ministerio de Educación, (Editorial Pueblo y Educación, Cuba, 2002) pp. 80-101.

[12] Serra, R., y otros, *La utilización del holograma como medio de enseñanza y de educación social en cuba como resultado del vínculo Universidad – Tecnología – Innovación*, Memorias del evento Universidad 2008, Cuba, (2008).

[13] UNESCO, Informe General, 2002, disponible en www.unesco.org/education. Consultado el 07 de octubre de 2009.

[14] Proyecto EDUTECH de Tecnologías y Educación Superior, 2002, disponible en www.edutech.ch/edutech. Consultado el 07 de octubre de 2009.

[15] Delicio, M., *Sistema de Actividades Educativas para promover el interés y la satisfacción de los alumnos de la enseñanza primaria por el Museo de Ciencia y Técnica de la Universidad Federal de Ouro Preto*, Tesis Doctoral, Cuba, (2003).

[16] Rustin, M. *Reason and unreason*, Psychoanalysis, science and politics, London, Continuum, 201-224, (2001).

[17] Salzberger, I., Henry, I., and Osborn, E., *The emotional experience of learning and teaching*, London, 23-52, (1983).

[18] Vigotsky, L., *Historia de las funciones psíquicas superiores*, (Editorial Científico-Técnica, La Habana, 1987).

[19] Vigotsky, L., *Pensamiento y Lenguaje*, (Edición Revolucionaria, La Habana, 1966).

[20] Vigostky, L., *El desarrollo de los procesos psicológicos superiores*, (Crítica, Barcelona, 1979).

[21] Galperin, P. Ya., *Sobre el método de formación por etapas de las acciones intelectuales*, Antología de la Psicología Pedagógica, (Pueblo y Educación, La Habana, Cuba, 1982).

[22] Talizina, N. F., *Psicología de la Enseñanza*, (Editorial Progreso, URSS, 1988).

[23] Iovanovich, M. L., *El pensamiento de Paulo Freire: sus contribuciones para la educación*. Acceso al texto completo: http://bibliotecavirtual.clacso.org.ar/ar/libros/freire/iovanovich.pdf, Consultado el 07 de octubre de 2009.

[24] Freire, P., *Pedagogia da Autonomia – Saberes necessários à prática educativa*, (Ed. Paz e Terra, 30ª ed., Brasil, 2004).

[25] González, V., *Teoría y metodología del uso de la televisión en circuito cerrado como parte del sistema de medios de enseñanza de la educación superior*, Tesis Doctoral, Cuba, (1984).

[26] Baglietto, M., Trabajo artístico disponible en www.**mireyabaglietto**.com, Consultado el 07 de octubre de 2009.

[27] Lunazzi, J. J., *DVD con Experimentos didácticos de Física*, disponible en www.ifi.unicamp.br/vie, Consultado el 07 de octubre de 2009.

[28] Lunazzi, J. J., *A luz congelada*, Universidade Estadual de Campinas, Revista Ciência Hoje **3**, 36-46 (1985).